\documentclass[twocolumn]{emulateapj}

\usepackage{graphicx}
\usepackage{epsfig}

\shorttitle{TESTING THE NO-HAIR THEOREM: II. BLACK-HOLE IMAGES}
\shortauthors{JOHANNSEN \& PSALTIS}

\begin{document}

%% LaTeX will automatically break titles if they run longer than
%% one line. However, you may use \\ to force a line break if
%% you desire.

\title{TESTING THE NO-HAIR THEOREM WITH OBSERVATIONS IN THE ELECTROMAGNETIC SPECTRUM:\\ II. BLACK-HOLE IMAGES}

\author{Tim Johannsen\altaffilmark{1} and Dimitrios Psaltis\altaffilmark{2}}
\affil{$^1$Physics Department, University of Arizona, 1118 E. 4th Street, Tucson, AZ 85721, USA; timj@physics.arizona.edu}
\affil{$^2$Astronomy Department, University of Arizona, 933 N.\ Cherry Ave., Tucson, AZ 85721, USA; dpsaltis@email.arizona.edu}

\begin{abstract}

According to the no-hair theorem, all astrophysical black holes are fully described by their masses and spins. This theorem can be tested observationally by measuring (at least) three different multipole moments of the spacetimes of black holes. In this paper, we analyze images of black holes within a framework that allows us to calculate observables in the electromagnetic spectrum as a function of the mass, spin, and, independently, the quadrupole moment of a black hole. We show that a deviation of the quadrupole moment from the expected Kerr value leads to images of black holes that are either prolate or oblate depending on the sign and magnitude of the deviation. In addition, there is a ring-like structure around the black-hole shadow with a diameter of $\sim$10 black-hole masses that is substantially brighter than the image of the underlying accretion flow and that is independent of the astrophysical details of accretion flow models. We show that the shape of this ring depends directly on the mass, spin, and quadrupole moment of the black hole and can be used for an independent measurement of all three parameters. In particular, we demonstrate that this ring is highly circular for a Kerr black hole with a spin $a\lesssim0.9M$, independent of the observer's inclination, but becomes elliptical and asymmetric if the no-hair theorem is violated. Near-future very-long baseline interferometric observations of Sgr A* will image this ring and may allow for an observational test of the no-hair theorem.

\end{abstract}

%Uncomment for PACS numbers title message
%\pacs{00.00, 20.00, 42.10}
% Keywords required only for MST, PB, PMB, PM, JOA, JOB? 
%\vspace{2pc}
%\noindent{\it Keywords}:
% Uncomment for Submitted to journal title message
%\submitto{\JPA}
% Comment out if separate title page not required
%\maketitle

\section{INTRODUCTION}

According to the no-hair theorem, the external spacetimes of black
holes are uniquely characterized by their masses and spins (Israel
1967, 1968; Carter 1971, 1973; Hawking 1972; Robinson 1975). This
theorem requires that the cosmic censorship conjecture (Penrose 1969)
holds and that the exterior spacetime is free of closed time-like
curves. Under these assumptions, all astrophysical black holes should
be fully described by the Kerr metric (Kerr 1963).

It is widely accepted that the universe contains an abundance of black
holes as inferred from the observations of the centers of nearby
galaxies (e.g., Tremaine et al.\ 2002), of our own galactic
center (Ghez et al. 2008; Gillessen et al. 2009), and of many galactic
binaries (e.g., McClintock \& Remillard 2006). Nonetheless, the
factual existence of an event horizon is yet unproven and has only
been inferred indirectly (e.g., Narayan, Garcia, \& McClintock 1997,
2001; Narayan \& Heyl 2002; McClintock, Narayan, \& Rybicki 2004; see
also Psaltis 2006). Alternative explanations for the nature of these
objects have been suggested, which include naked singularities (Manko
\& Novikov 1992), exotic stellar objects (Friedberg, Lee, \& Pang
1987; Mazur \& Mottola 2001; Barcel\'{o} et al. 2008), as well as a
breakdown of general relativity itself on horizon scales (e.g., Yunes
\& Pretorius 2009; c.f. Psaltis et al. 2008).

A test of the no-hair theorem can both identify the observed dark
compact objects with Kerr black holes and verify the validity of
general relativity in the strong-field regime. Indeed, within general
relativity, if a compact object is not a Kerr black hole, then its
external spacetime will not satisfy the no-hair
theorem. Alternatively, if general relativity is not valid in the
strong-field regime, the external spacetime of a compact object that
is surrounded by a horizon may violate the no-hair theorem (see,
however, Psaltis et al.\ 2008).  Mass and spin are the first two
multipole moments of a black-hole spacetime. If the no-hair theorem is
correct, then all higher multipole moments only depend on the mass and
spin, and any deviation from the Kerr moments has to be
zero. Consequently, the no-hair theorem can be tested by measuring (at
least) three multipole moments of such a spacetime (Ryan 1995).

In part I of this series of papers (Johannsen \& Psaltis 2010;
  hereafter Paper I), we investigated a framework for testing the
no-hair theorem with observations of compact objects in the
electromagnetic spectrum. Based on a quasi-Kerr spacetime that
contains an independent quadrupole moment (Glampedakis \& Babak 2006),
we analyzed in detail the spacetime properties that are critical for
such observations as a function of the mass, spin, and quadrupole
moment. We showed that already very moderate changes of the quadrupole
moment lead to significant alterations of various quantities that
determine observables. In particular, we explored the effect of
changing the quadrupole moment on the locations of the innermost
stable circular orbit (ISCO) and of the circular photon orbit, as well
as on the lensing and redshift of photons.

There has been, already, substantial work on potential tests of the
no-hair theorem with observations of gravitational waves from extreme
mass-ratio inspirals (Ryan 1995, 1997a, 1997b; Barack \& Cutler 2004, 2007; Collins \& Hughes 2004;
Glampedakis \& Babak 2006; Gair et al.\ 2008; Li \& Lovelace 2008; Apostolatos et
al.\ 2009; Vigeland \& Hughes 2010).  In this series of papers, we
show that observations of black holes in the electromagnetic spectrum
may also allow for a clean test of the no-hair theorem. In
particular, we identify different observables that probe the
quadrupole moments of the spacetimes but depend very weakly on the
usual astrophysical complications, such as the flow geometry, the mode
of emission, and the variability of the accretion flows that generate
the photons we detect from black holes.

Imaging observations of accreting black holes at (sub$-$)mm wavelengths
using very-long baseline interferometry (VLBI) promise to enable
unprecedented views of the vicinities of black-hole horizons. Recent
VLBI-observations along only three baselines resolved Sgr~A$^*$, the
black hole in the center of the Milky Way, on a scale comparable to
its event horizon and provided evidence for sub-horizon scale
structures (Doeleman et al.\ 2008) as well as for the presence of an event 
horizon (Broderick, Loeb, \& Narayan 2009). Far greater resolution can be
achieved by adding either existing or planned telescopes located at
various places on the Earth (Fish \& Doeleman 2009). The black hole in
the center of M87 (Broderick \& Loeb 2009) as well as a small number
of other nearby supermassive black holes (Psaltis 2008) offer
additional targets for horizon-scale imaging that can be utilized in
the near future.

Imaging observations are expected to be able to resolve the shadows of
black holes and lead to the determination of their spins and
inclinations (Falcke et al. 2000; Broderick \& Loeb 2005, 2006; Fish
\& Doeleman 2009). Templates for images of accretion flows around Kerr
black holes within general relativity that are suitable to these
observations have been reported by a number of authors (Bardeen 1973;
Speith et al. 1995; Fanton et al. 1997; Falcke et al. 2000; Takahashi
2004; Beckwith \& Done 2004, 2005; Dexter \& Agol 2009; Broderick \&
Loeb 2005, 2006; Yuan et al. 2009). Recently, Bambi \& Freese (2009)
explored the possibility of using black-hole images to test whether
black holes violate the Kerr bound $a\le M$.

In this paper, we study the properties of the images of compact
objects that violate the no-hair theorem using the quasi-Kerr
formalism we developed in Paper I.  We calculate numerically the
mapping between locations in the vicinity of a black hole and
positions in the observer's sky using the mass, spin, and quadrupole
moment of the spacetime as independent parameters. We investigate the
impact of varying the quadrupole moment on the properties of this
mapping and show that the images of the accretion flows around compact
objects that violate the no-hair theorem are expected to have
prolate or oblate geometries.

Measuring the spacetime moments from the images of an accretion flow
will be, of course, very model dependent and limited by our lack of
understanding of the intrinsic geometry of the flow itself. For
example, prolate images of the inner accretion may be the result of
resolving the formation region of a jet and not of a violation of the
no-hair theorem (see, e.g., Broderick \& Loeb 2009). Moreover, a measurement of the spin from an image of the shadow alone is difficult (e.g., Falcke et al. 2000; Takahashi 2004) and might require complementary observations such as a multiwavelength study of polarization (Broderick \& Loeb 2006; see also Schnittman \& Krolik 2009, 2010).

Additionally, accretion flows are very turbulent and variable at timescales much
shorter than the rotation period of the Earth, which sets the
characteristic integration time for an interferometric imaging
observation. If this time variability is produced by a highly
coherent, orbiting inhomogeneity in the accretion flow, it may allow
measuring the properties of the compact object via non-imaging
techniques (Doeleman et al.\ 2009). The variability, however, will
limit and may prohibit altogether the ability of obtaining a clean
image of the accretion flow.

The images of optically thin accretion flows around black holes,
however, reveal a characteristic bright ring at the projected radius
of the circular photon orbit along null geodesics (Beckwith \& Done 2005) with properties that
remain constant even as the underlying accretion flows are highly
variable. This bright ring is the result of the light rays that orbit
around the black hole many times before they reach the distant
observer, and, therefore, have a much larger path length through the
optically thin accretion flow. These photons can make a significant contribution to the total disk emission and produce higher-order images (Cunningham 1976; Laor, Netzer, \& Piran 1990; Viergutz 1993; Bao, Hadrava, \& {\O}stgaard 1994; ${\rm \check{C}ade\check{z}}$, Fanton, \& Calvani 1998; Agol \& Krolik 2000; Beckwith \& Done 2005).

We use our formalism to show that the bright emission ring is circular
for a Schwarzschild black hole and remains nearly circular for Kerr
black holes. On the other hand, if the quadrupole moment is left as an
independent parameter, the ring shape changes significantly and becomes
asymmetric. The degree of asymmetry is a direct measure of the
violation of the no-hair theorem. We show that the diameter of the ring depends only very weakly on the spin and quadrupole moment of the black hole and can be used to directly measure the mass of the object. In addition, the ring is displaced off center in the image plane in the case of rotating black holes (Beckwith \& Done 2005; see, also, Takahashi 2004), and we show that the displacement is a direct measure of the object's spin, modulo the disk inclination.

In Section~2 we briefly review the framework for testing the no-hair
theorem with observations in the electromagnetic spectrum. We simulate
images of black holes and of the photon rings in Sections~3 and 4,
respectively, and show how they depend on the mass, spin, and
quadrupole moment of a given black hole. In Section~5 we quantify these dependencies before we
discuss our results and their implications for future observations in Section~6.

\section{TESTING THE NO-HAIR THEOREM WITH BLACK-HOLE IMAGES}

The no-hair theorem establishes the claim that astrophysical black
holes are uniquely characterized by their mass $M$ and spin $J$, i.e.,
by only the first two multipole moments of their exterior spacetimes
(Israel 1967, 1968; Carter 1971, 1973; Hawking 1972; Robinson
1975). As a consequence of the no-hair theorem, all higher order
moments are already fully determined and obey the relation (Geroch
1970; Hansen 1974)
\begin{equation}
M_{l}+{\rm i}S_{l}=M({\rm i}a)^{l}.
\label{kerrmult}
\end{equation}
Here, $a\equiv J/M$ is the spin parameter, and the multipole moments are
written as a set of mass multipole moments $M_l$ which are nonzero for
even values of $l$ and as a set of current multipole moments $S_l$
which are nonzero for odd values of $l$. The zeroth order mass moment
is equal to the mass of the black hole, $M_0=M$, whereas the first order
current moment is its angular momentum, $S_1=J$.

This theorem is based on two technical assumptions. First, any spacetime singularity must be enclosed by an event
horizon (the cosmic censorship conjecture, Penrose 1969), and second,
 there exist no closed timelike loops in the exterior domain of
the black hole.

The no-hair theorem can be tested by measuring (at least) three
different multipole moments, which have to be related by expression
(\ref{kerrmult}) if this theorem is correct (Ryan 1995). In Paper~I,
we investigated a framework for extracting three multipole moments of
a black-hole spacetime with observations in the electromagnetic
spectrum. We used a quasi-Kerr metric (Glampedakis \& Babak 2006),
which incorporates an independent quadrupole moment and parameterizes
a potential deviation from the Kerr quadrupole in terms of the
parameter $\epsilon$, i.e.,
\begin{equation}
Q=-M\left(a^2+\epsilon M^2\right).
\label{qradmoment}
\end{equation}
This reduces smoothly to the Kerr quadrupole moment in the limit
$\epsilon\rightarrow0$ in accordance with relation (\ref{kerrmult}).

If the no-hair theorem is valid, then $\epsilon=0$. If, however, a
nonzero value of the parameter $\epsilon$ is measured, then the
compact object cannot be a general-relativistic black hole. Within
general relativity, it may be a different type of star or an exotic
configuration of matter (see Collins \& Hughes 2004; Hughes 2006). On
the other hand, if the compact object is otherwise known to possess an
event horizon and a regular spacetime, then a nonzero value of the
parameter $\epsilon$ implies that the no-hair theorem is incorrect and
general relativity does not accurately describe the near-horizon
spacetimes of black holes.

In Paper I, we showed that the observable properties of this
quasi-Kerr spacetime depend significantly on the parameter $\epsilon$
in the vicinity of the black hole. In particular, we demonstrated that
the radius of the ISCO increases by $\sim$20\% for a value of the
quadrupolar correction $\epsilon=0.5$ at a spin $a=0.4M$. Since
the ISCO marks the inner edge of the accretion disk, it critically
impacts the high-energy part of the emitted disk
spectrum. Furthermore, we showed that the circular photon orbit
experiences a shift of similar magnitude and that the observed
redshift of a photon emitted by a particle on the ISCO decreases by
$\sim$25\% for values of the quadrupolar parameter $\epsilon=1.0$ and
spin $a=0.4M$, respectively. In addition, we demonstrated the
effects of changing the quadrupole moment on the strong lensing
experienced by photons in the neighborhood of the black hole.

For nonzero values of the parameter $\epsilon$, the quasi-Kerr metric
is a solution of the Einstein equations up to the quadrupole order
and, therefore, can only be used for values of the spin $a\lesssim0.4M$ and of the radius larger than a cutoff that depends on both the
spin and the quadrupole moment. For reasonable perturbations of the
quadrupole, however, the cutoff always lies inside of the circular
photon orbit and plays only a very minor role in our analysis. The
requirement that the spin is not near the maximal Kerr value allows us
to apply this method to images of Sgr~A$^{*}$, for which first, although uncertain
estimates from millimeter VLBI observations indicate that the black
hole is not spinning rapidly (Broderick et al. 2009).

\section{THE APPARENT SHAPE OF QUASI-KERR BLACK HOLES}

\label{shapes}

In this section, we calculate numerically the mapping between
  different locations in the accretion flow around a quasi-Kerr black
  hole and the observer's sky and discuss its characteristics. This
mapping depends significantly on the value of the parameter
$\epsilon$ due to the modifications in the spacetime properties
discussed in Paper~I, such as their effects on the light bending and
  redshift experienced by photons, as well as on the location of the
ISCO and the photon orbit.

\begin{figure}
\begin{center}
\includegraphics[width=0.4\textwidth]{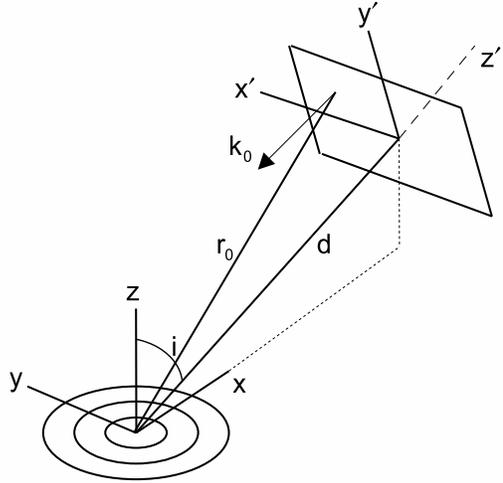}
\end{center}
\caption{The geometry we used for simulating black-hole images,
    as projected on a Euclidean space. The black hole is located at
  the origin of the Cartesian coordinate system $(x,y,z)$. The image
  plane of a distant observer is located at a distance $d$ from the
  black hole, at an inclination angle $i$, and with its center on
    the $x$--$z$ plane. Photon trajectories are integrated backwards
  starting on the image plane at coordinates $(x',y')$ with uniform
  initial  momentum vector $\vec{k}_0=-k_0\hat{z}'$ at a distance
  $r_0$ from the black hole.}
\label{geometry}
\end{figure}

We developed an algorithm that maps any initial configuration of
photons around the black hole into the plane of the sky viewed by a
distant observer. We integrate the full second-order geodesic
equations in the quasi-Kerr spacetime via a fourth-order Runge-Kutta
method with adaptive stepsize.  Figure~\ref{geometry} shows the
geometry we use. The image plane is located at a distance $d$ away
from the black hole, at an inclination angle $i$, and with its center
on the $x$-$z$ plane. We integrate photon trajectories backwards
starting on the image plane and terminating at the surface of last
scattering in the accretion flow. Modeling the thermodynamic
properties of the accretion flow and integrating the radiative
transfer equation is beyond the scope of this paper. In this section,
we assume for simplicity that the last scattering surface of the
photons is in the equatorial plane around the black hole, on the
surface of an infinitesimally thin accretion disk.  In the next
section, we will discuss the situation in which the accretion flow is
optically thin, as is expected to be the case for Sgr~A$^{*}$ at
sub-mm wavelengths.

As an initial condition, we distribute the photons in a square grid on
the image plane with a spacing of $\Delta r^\prime=0.025M$. We set
their initial 3-momentum vectors $\vec{k}_0$ to be perpendicular to the image
plane and uniform (see Appendix~A). In order to visualize the mapping
between the plane of the accretion disk and the observer's sky we
consider a set of concentric equatorial target rings that extend from
$r=2.6M$ to $r=7.6M$ in steps of $1M$. Since the
images are scale-invariant with respect to the mass of the black hole,
we express all physical quantities in units of mass. In all
calculations we have imposed a cutoff at $r=2.6M$ in
accordance with the range of validity of the metric (see Paper~I), and
we terminate the integration if a photon enters the excluded region
$r<2.6M$. We plot on the image plane only those photons that
reach one of the target rings in the accretion disk.

\begin{figure*}[t]
\begin{center}
\psfig{figure=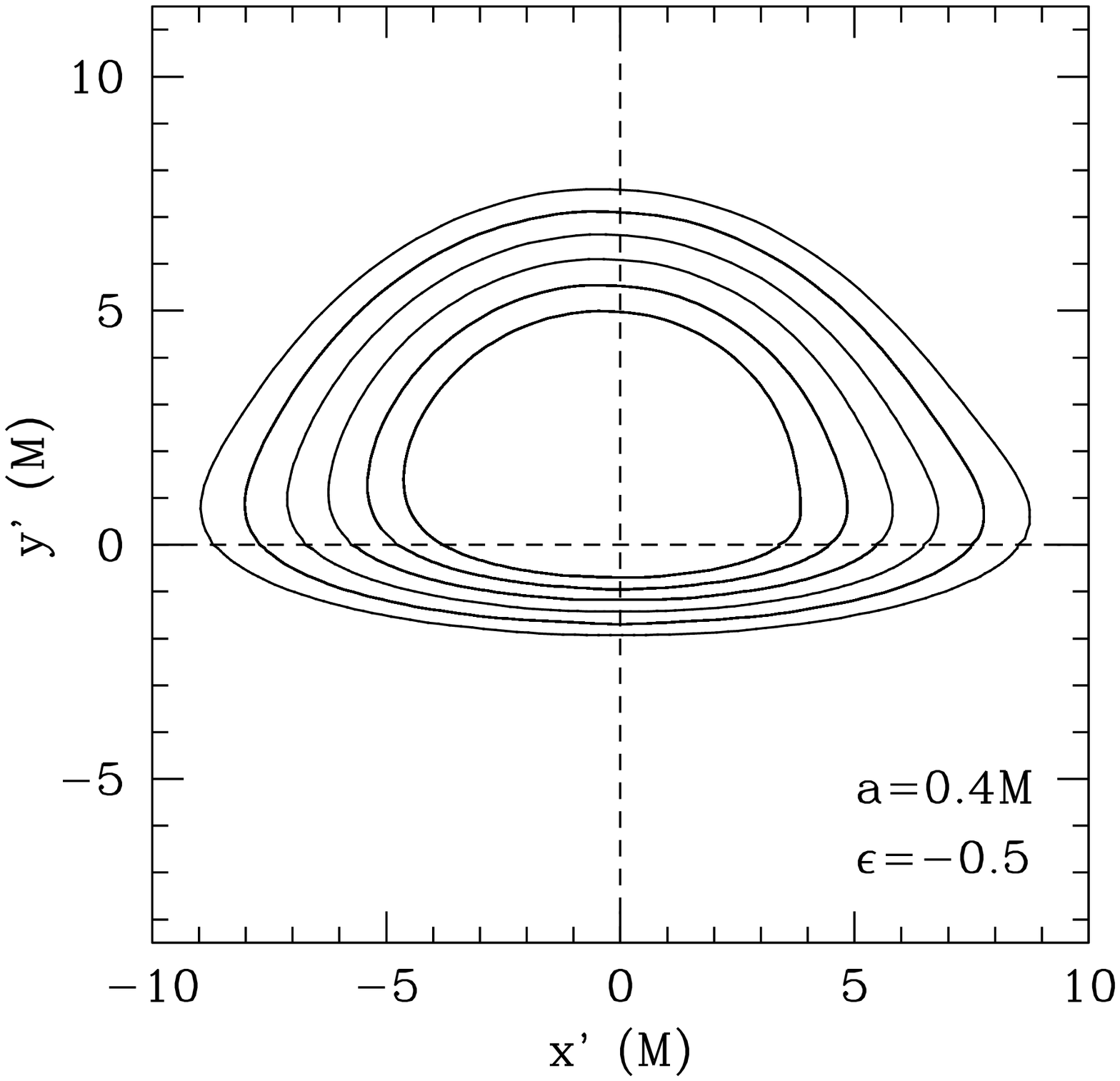,height=2.1in}
\psfig{figure=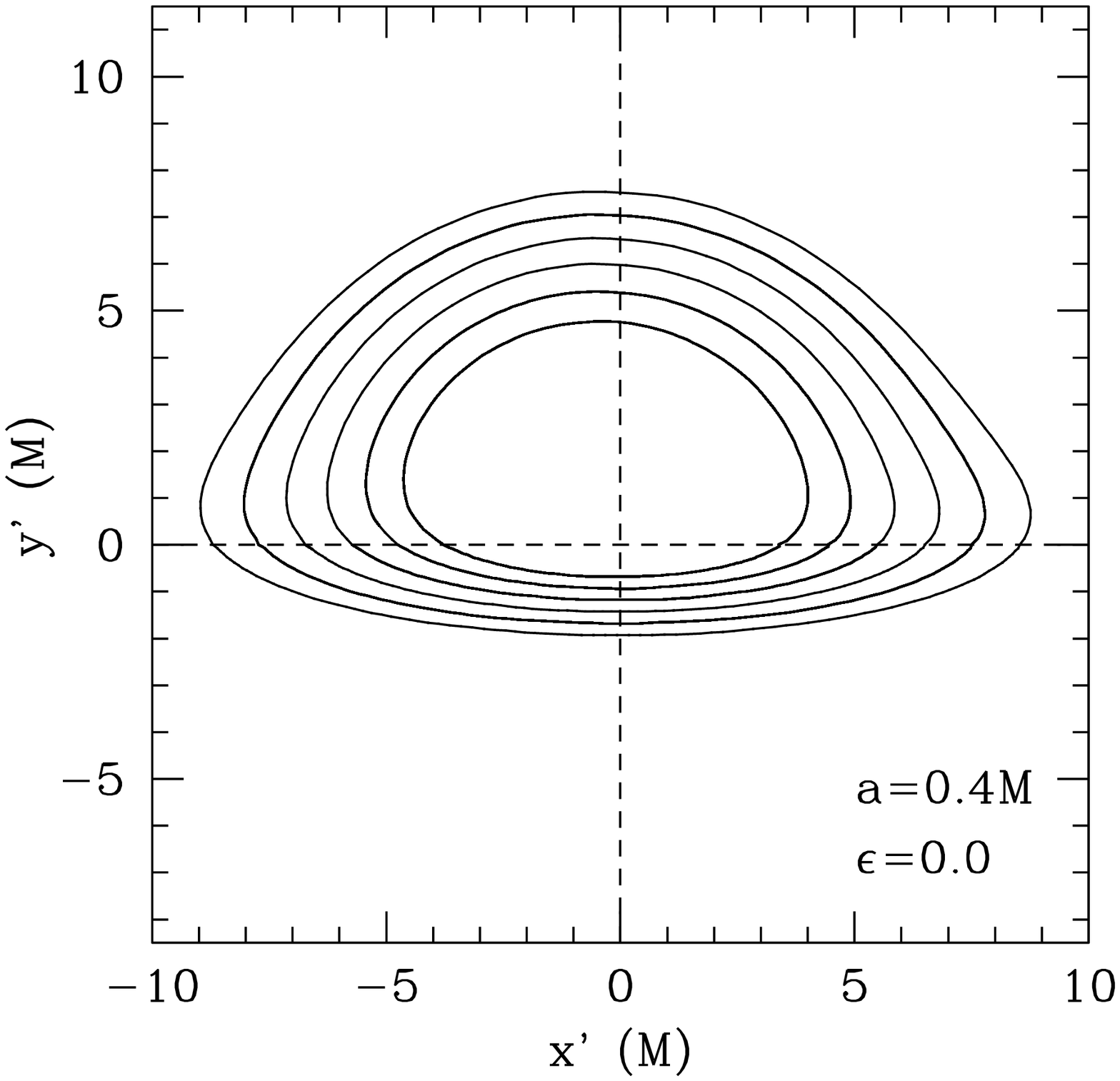,height=2.1in}
\psfig{figure=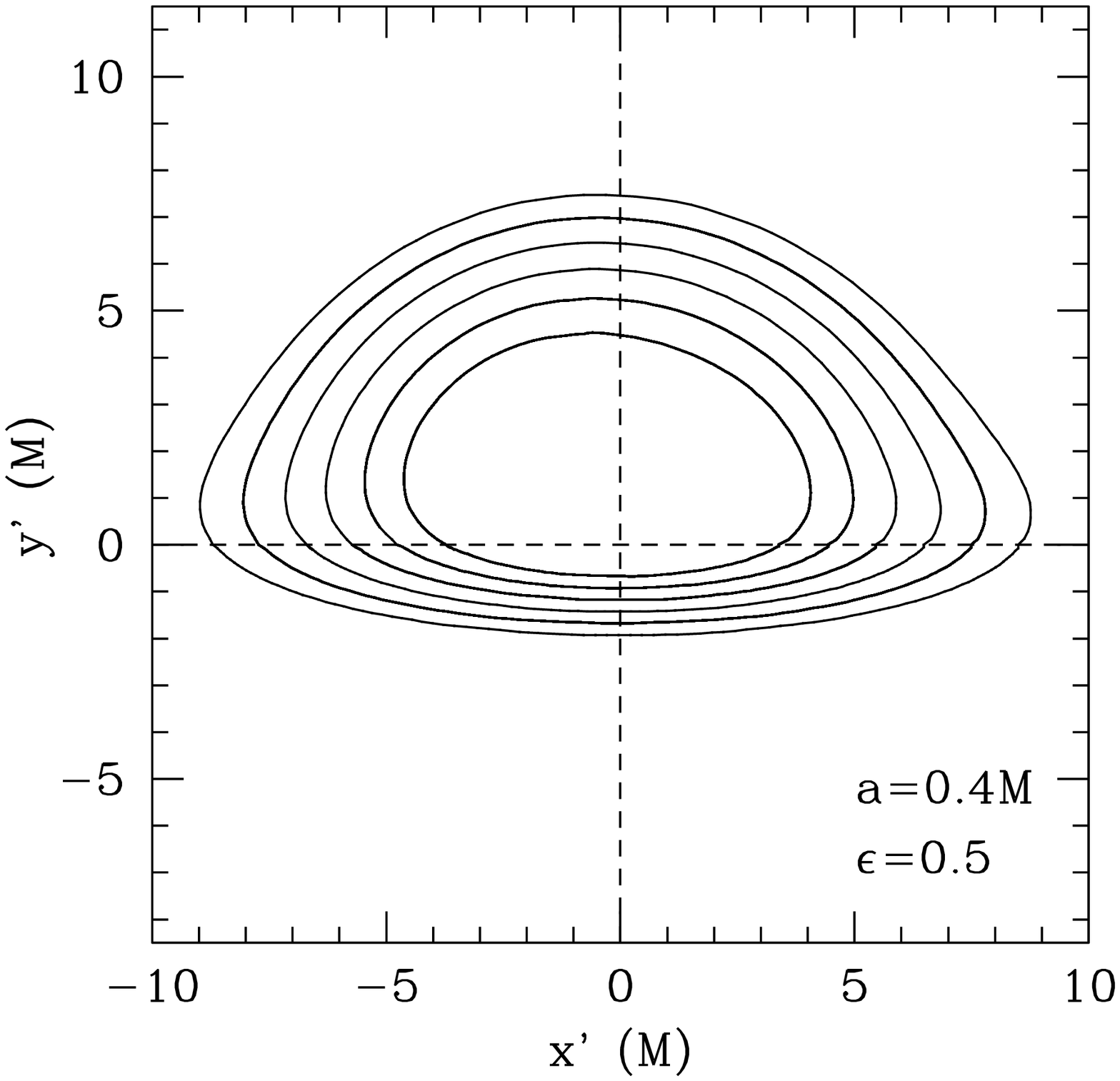,height=2.1in}
\end{center}
\caption{Concentric rings on the equatorial plane around a black hole,
  as seen by a distant observer located at an inclination of $\cos
  i=0.25$. The spacetime of the compact object has a spin of
  $a=0.4M$ and a quadrupolar correction parameter ({\em left})
  $\epsilon=-0.5$, ({\em center}) $\epsilon=0.0$, and ({\em right})
  $\epsilon=0.5$. In all cases, frame dragging shifts the center of
  the images to the left. The effect of the quadrupolar correction is
  most apparent in the shape of the innermost ring. For
  $\epsilon=-0.5$, the innermost ring is more prolate compared to the
  Kerr black hole, while for $\epsilon=0.5$ it is more oblate.}
\label{BHimages}
\end{figure*}

In Figure~\ref{BHimages} we plot images of black holes with spin
$a=0.4M$ viewed from a distant observer at an inclination angle
$\cos i=0.25$ for values of the quadrupolar parameter (from left to
right) $\epsilon=-0.5,~0.0,~0.5$.  The images consist of the
projection of the target rings onto the observer's sky. The rings are
distorted by light bending. Gravitational redshift, Doppler boosting,
and beaming play no role in these images, because we have not
  incorporated in our calculation the evolution of the radiative
  intensity.  In all images, the centers of the deformed rings are
shifted to the left due to frame dragging. 

\begin{figure}[t]
\begin{center}
\psfig{figure=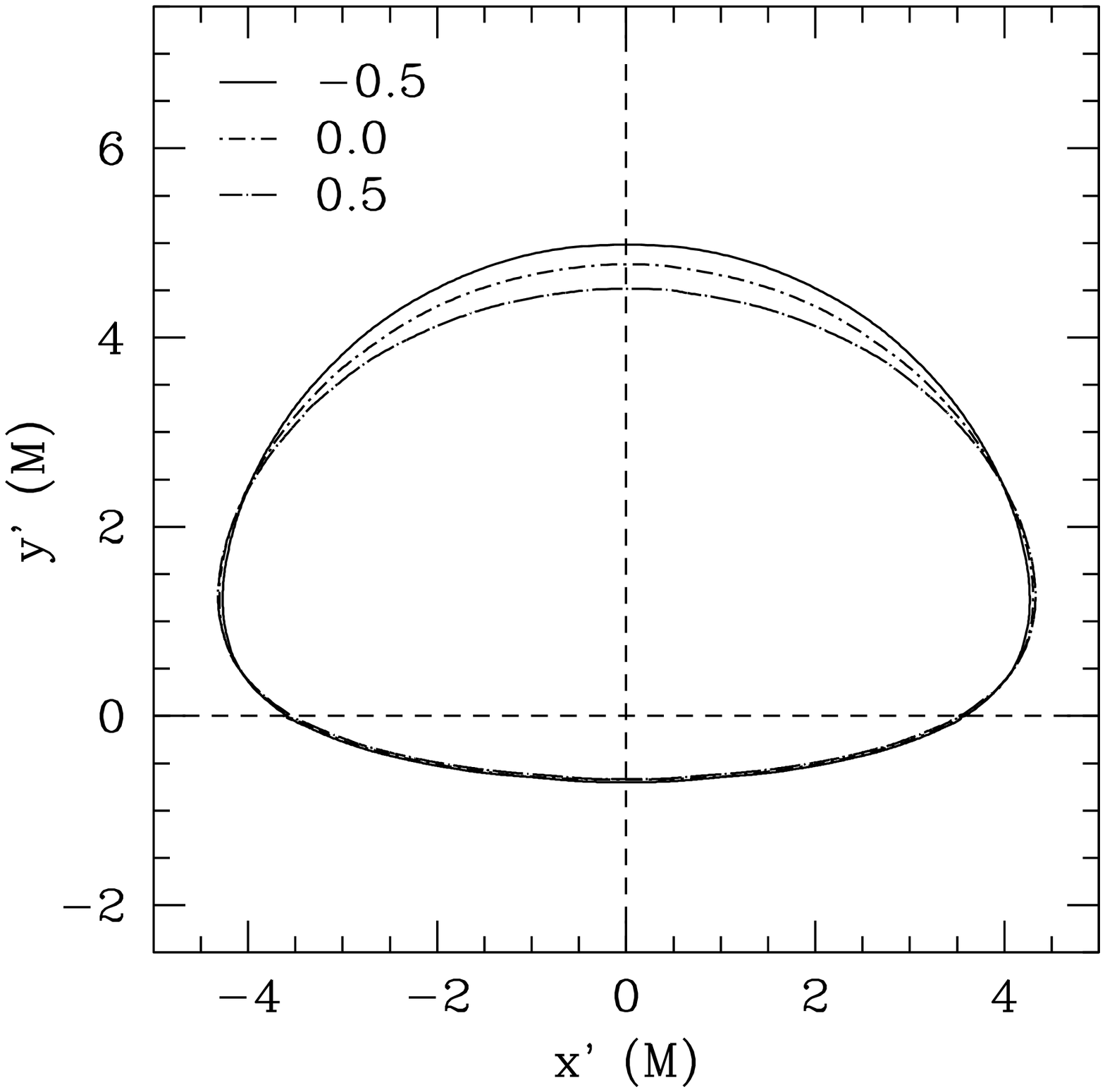,height=2.5in}
\psfig{figure=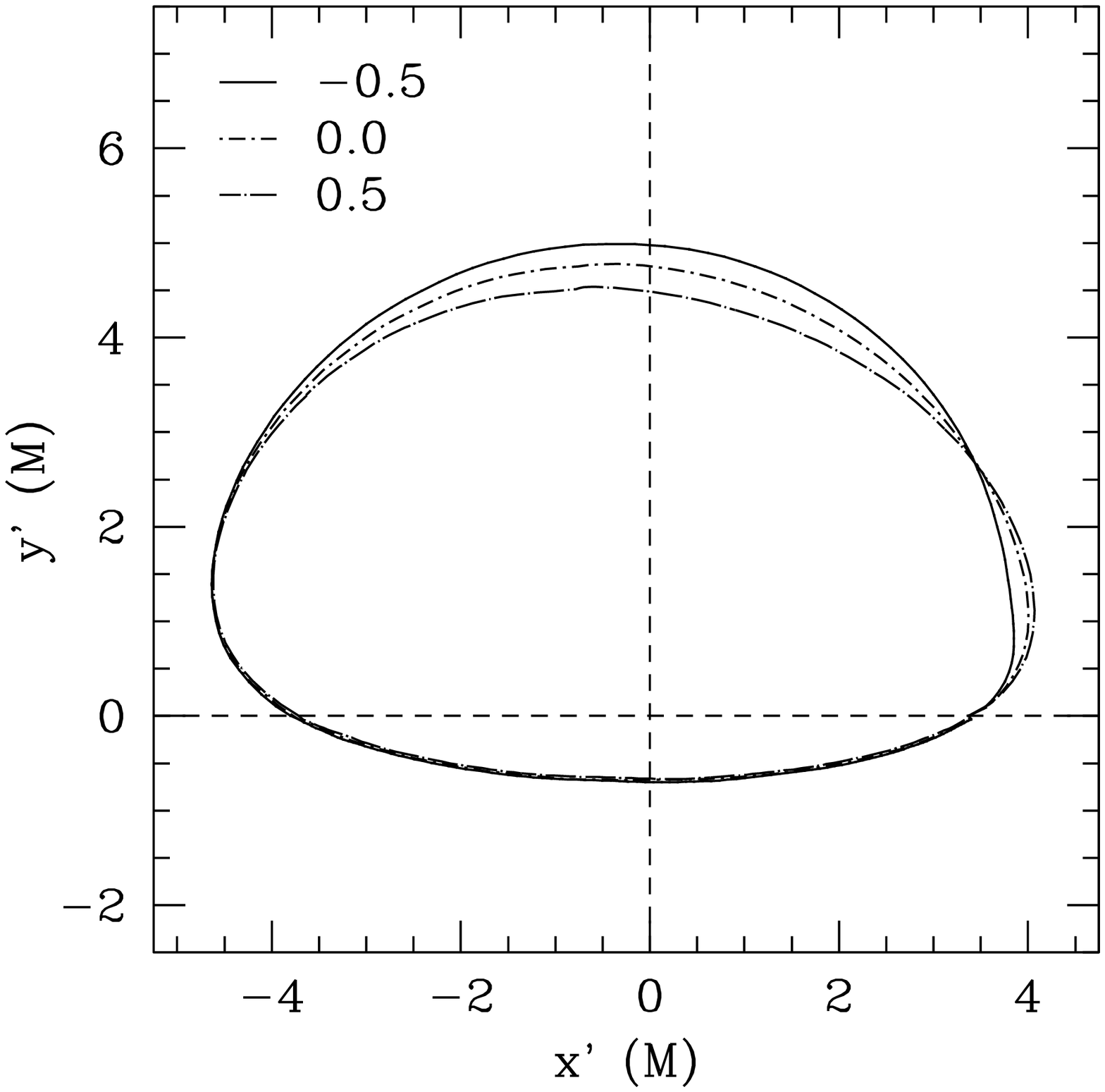,height=2.5in}
\end{center}
\caption{Images of the innermost rings shown in Figure~\ref{BHimages}
  for three values of the parameter $\epsilon$.  The top panel
  corresponds to a Schwarzschild black hole, whereas the bottom panel
  corresponds to a Kerr black hole with $a=0.4M$. The
  inclination angle is set to $\cos i=0.25$ in both cases.}
\label{r2}
\end{figure}

The effect of changing the quadrupole moment on the images is
only apparent at small radii. In Figure~\ref{r2} we show explicitly
how violating the no-hair theorem alters the images of the innermost
rings we considered, i.e., the ones at $r=2.6M$, for a
Schwarzschild black hole and a Kerr black hole with $a=0.4M$. In
both cases, we set the inclination to $\cos i=0.25$ and display the
innermost ring for three values of the parameter
$\epsilon=-0.5,~0.0,~0.5$.  For the perturbed Schwarzschild black
hole, increasing the value of the parameter $\epsilon$ makes the
projected ring more oblate, because of the increased amount of
gravitational lensing experienced by the photons (see Paper~I).  In
the case of nonzero spin, a change in the quadrupole moment also
affects the left-right asymmetry of the image and especially its right
edge that corresponds to the receding part of the accretion
disk. Increasing the value of the parameter $\epsilon$ leads to a
wider image.

\section{PHOTON RINGS}

The dependence of the black-hole images on their quadrupole moments
that we investigated in the previous section allows us, in principle,
to use imaging observations to map the properties of their spacetimes.
In practice, however, the change in the oblateness of even the closest
rings to the black-hole horizons is very modest and can be masked by
anisotropies in the emission from the accretion flow and its
variability. Unless an orbiting density inhomogeneity (a ``hot spot'')
can be securely identified and imaged throughout its orbit (e.g.,
Broderick \& Loeb 2005), time-averaged overall images of black-hole
accretion disks may appear {\em a priori\/} not well suited for
testing the no-hair theorem. There is, however, an observable
structure in the images of optically thin accretion flows around black
holes that suffers only marginally from astrophysical complications
and carries very strong signatures of the quadrupole moments of the
underlying spacetimes.

\begin{figure}
\begin{center}
\includegraphics[width=0.4\textwidth]{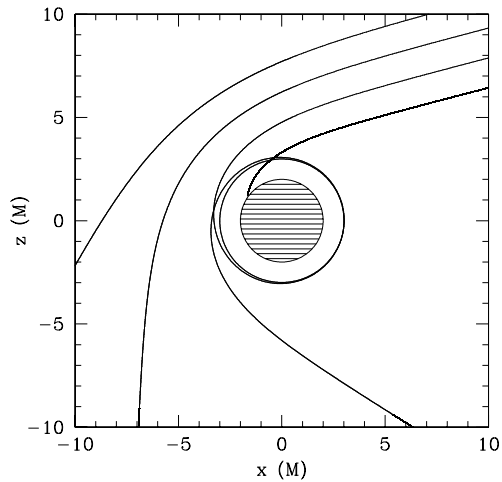}
\end{center}
\caption{Light rays around a Schwarzschild black hole illustrating the
  emergence of the bright emission ring. Several light rays approach
  the black hole from the top right corner. If a ray reaches the
  photon ring with a 3-momentum that is nearly tangential to the photon
  orbit, it orbits around the black hole several times, while all
  other rays are either immediately scattered or captured by the black
  hole. The footpoints of the orbiting light rays on the image plane
  will be brighter than those of the nearby rays. The shaded region
  marks the event horizon.}
\label{ringorbits}
\end{figure}

Consider a geometrically thick accretion flow imaged at a wavelength
at which the emission is optically thin; this is the case for
Sgr~A$^{*}$ at sub-mm wavelengths. The brightness of the image at any
given point in the observer's sky will depend on the length of the
optical path along the corresponding light ray that passes through the
region of high emissivity in the accretion flow. Most light rays that
originate from the image plane either emerge on the far side of the
black hole, after experiencing gravitational bending, or intercept the
black-hole horizon. A small set of light rays, however, that approach
the event horizon orbit closely around the black hole several times
before they escape towards the far side and can, therefore, make a significant contribution to the total flux (Cunningham 1976; Laor, Netzer, \& Piran 1990; Viergutz 1993; Bao, Hadrava, \& {\O}stgaard 1994; ${\rm \check{C}ade\check{z}}$, Fanton, \& Calvani 1998; Agol \& Krolik 2000; Beckwith \& Done 2005). This is illustrated in
Figure~\ref{ringorbits}, which shows the paths of a number of light
rays that approach the horizon of a black hole from the top right
corner at an inclination of $\cos i=0.25$ relative to the spin axis
$z$. These light rays orbit several times around the black-hole
horizon at nearly constant radius (see, also, Bardeen 1973).

The integral of the emissivity along these light rays is very large,
compared to that along nearby light rays, and results in a significant
increase in the brightness at their footpoints on the image plane. The
locus of the footpoints of these light rays on the image plane is a
ring. This leads to the emergence of a ring in images of optically
thin accretion flows that is substantially brighter than the
background and whose shape and position in the image plane depend on the spin of the black hole and on the inclination of the accretion disk (Beckwith \& Done 2005).  Such bright rings of emission are clearly visible in the
images of all time-dependent general relativistic simulations of
accretion flows reported to date (see, e.g., the right panel of Fig.~5
in Mo$\acute{\rm s}$cibrodzka et al. 2009; panels 1 and 3 in Fig.~1 in
Dexter, Agol, \& Fragile 2009; Fig.~1 in Shcherbakov \& Penna 2010).

In this section, we demonstrate that the bright rings of emission
(hereafter photon rings) carry an unmistakable signature of the
black-hole spacetime and, in particular, of its quadrupole moment. In
order to study the sizes and shapes of these rings in different
configurations, in a way that is independent of the physical
conditions in the accretion flow, we use the following working
definition of the position of the bright photon rings on the image
plane: we identify them with the locus of footpoints of those light
rays that leave the observer perpendicular to the image plane and cross
the equatorial plane of the black hole at least twice, before emerging
on the opposite side.

\begin{figure}
\begin{center}
\includegraphics[width=0.4\textwidth]{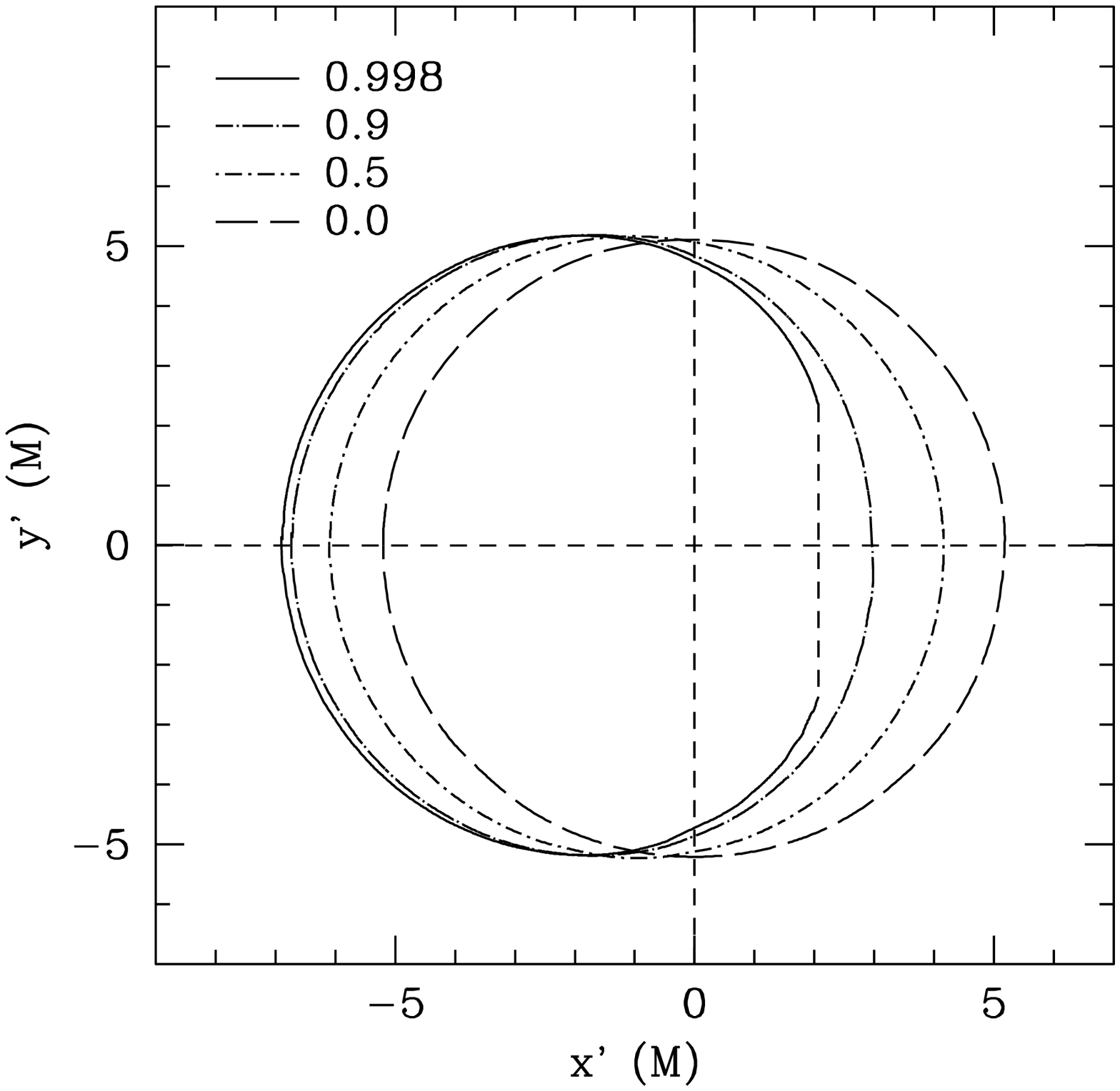}
\end{center}
\caption{The dependence of the bright photon ring seen by a distant
  observer on the spin of a Kerr black hole. Increasing the spin leads
  to a displacement of the photon ring with minimal deformation of its
  shape.  In all cases, the inclination of the observer corresponds to
  $\cos i=0.25$.  }
\label{ringkerr}
\end{figure}

We first consider the dependence of the position of the photon rings
on the spin of a Kerr black hole, i.e., of one that satisfies the
no-hair theorem. Figure~\ref{ringkerr} shows that increasing the spin
leads to a substantial displacement of the centroid of the photon ring
with respect to the geometric center of the spacetime. This
displacement is $\simeq1M$ for even moderate ($a=0.5M$) values of the
spin and can be as large as $\simeq2M$ for maximally rotating black
holes. The striking property of these rings, however, is the fact that
their shapes remain practically circular for values of the spin $a\lesssim0.9M$, even though the geometry of
the Kerr spacetime is highly non-spherically symmetric. Only at the
maximum spin, the receding part of the photon ring becomes
asymmetric. Even in this extreme case, however, the maximum difference
between the major and minor axes of the ring is only $\simeq1.5M$. Note
here that, for the case $a=0.998M$, the radius of the circular photon 
orbit almost coincides with the event horizon, which introduces a
 noticeable numerical error on the rightmost part of the ring; for this 
reason, we use a short-dashed line to represent this uncertainty.

\begin{figure}[t]
\begin{center}
\includegraphics[width=0.4\textwidth]{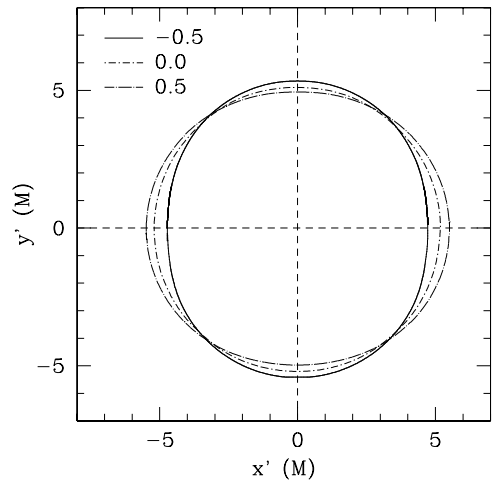}
\includegraphics[width=0.4\textwidth]{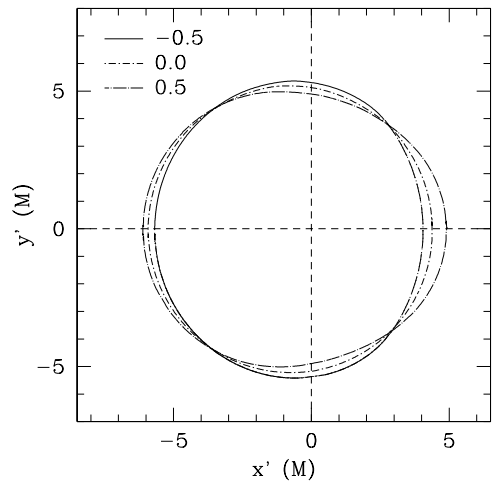}
\end{center}
\caption{The dependence of the bright photon rings on the degree of
  violation of the no-hair theorem, parametrized by the difference
  $\epsilon$ between the quadrupole moment of the spacetime and its
  Kerr value.  The top panel corresponds to a Schwarzschild black hole
  whereas the bottom panel corresponds to a Kerr black hole with
  $a=0.4M$. In both cases, the inclination of the observer is $\cos
  i=0.25$. Violation of the no-hair theorem causes the ring to
  become ellipsoidal, if the black hole is static, or even more
  asymmetric, if the black hole is rotating.}
\label{rings}
\end{figure}

The symmetry of the photon rings changes significantly if the no-hair
theorem is violated and the quadrupole moment of the spacetime takes on a
non-Kerr value. In Figure~\ref{rings}, we show the photon rings for a
(quasi$-$)Schwarzschild and a (quasi$-$)Kerr black hole with spin
$a=0.4M$ and for three different values of the quadrupolar
correction $\epsilon=-0.5,~0.0,~0.5$. As the degree of violation of
the no-hair theorem increases, the photon ring around a static black
hole becomes oblate or prolate (depending on the sign of the parameter
$\epsilon$), while its geometric center remains the same. In the case
of a spinning black hole, frame dragging introduces an additional
asymmetry between the approaching and the receding parts of the ring.
For the spinning black holes with a value of the quadrupolar parameter 
$\epsilon=-0.5$, the rightmost part of the ring corresponds to photons 
that propagate in a region of spacetime in which our perturbative 
approach breaks down; we, therefore, indicated this uncertainty using a 
long-dashed line.

In both cases shown in Figure~\ref{rings}, the difference between the
major and minor axes of the photon rings is $\simeq 1-2M$, even for
modest degrees of violation of the no-hair theorem. The degree of
asymmetry increases with increasing inclination and vanishes when the
black hole is viewed pole on.  The oblateness and asymmetry of the
bright photon rings in the images of accreting black holes carries,
therefore, a quantitative measure of the degree of violation of the
no-hair theorem, modulo the inclination of the observer. We systematically analyze the impact of the mass, spin, quadrupolar parameter, and disk inclination on the shape and position of the ring in the next section.

\section{PHOTON RING PROPERTIES}

In this section, we analyze in detail the dependence of the image of the photon ring for a given black hole on its mass $M$, spin $a$, disk inclination angle $i$, and quadrupolar correction parameter $\epsilon$.

In order to quantify the effect of changing this set of parameters on the shape and location of a photon ring in the image plane, we define the displacement and the asymmetry of the ring in the following way. First we define the horizontal displacement $D$ of the ring by the expression
\begin{equation}
D\equiv\frac{\left|x'_{\rm max}+x'_{\rm min}\right|}{2},
\end{equation}
where $x'_{\rm max}$ and $x'_{\rm min}$ are the maximum and minimum abscissae of the ring in the image plane, respectively. Due to the reflection symmetry across the equatorial plane there is no displacement in the vertical direction, whereas in the horizontal direction the displacement can be as large as $D\simeq2M$ for rapidly spinning black holes.

Next, we define the average radius $<R>$ of the ring by the expression
\begin{equation}
<R>\equiv\frac{\int_0^{2\pi}R{\rm d}\alpha}{\int_0^{2\pi}{\rm d}\alpha},
\end{equation}
where
\begin{equation}
R\equiv\sqrt{(x'-D)^2 + y'^2}
\end{equation}
and 
\begin{equation}
\tan\alpha=\frac{y'}{x'}.
\end{equation}
We also define the ring diameter $L$ by the expression
\begin{equation}
L\equiv2<R>.
\end{equation}

Finally, we define the asymmetry $A$ of the ring image by the expression
\begin{equation}
A\equiv 2 \sqrt{ \frac{\int_0^{2\pi}(R-<R>)^2{\rm d}\alpha}{\int_0^{2\pi}{\rm d}\alpha} }.
\end{equation}

\begin{figure}[t]
\begin{center}
\includegraphics[width=0.4\textwidth]{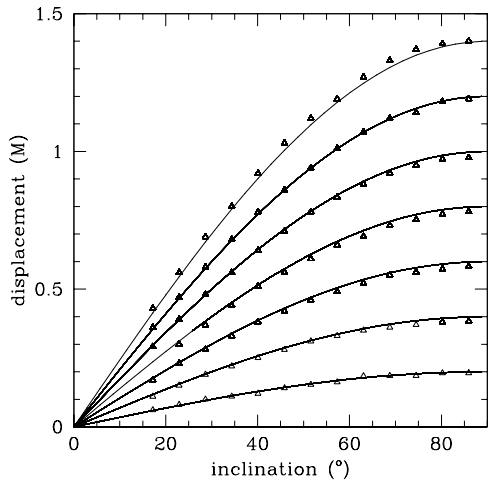}
\end{center}
\caption{The ring displacement versus the disk inclination for a Kerr black hole with spins (top to bottom) $a=0.7M,~0.6M,...,0.1M$. The solid lines show the functional form (\ref{displacementformula}) with the parameter $\epsilon$ set to zero.}
\label{displacementfits}
\end{figure}

We simulated ring images for various values of the spin $a$, inclination $i$, and the parameter $\epsilon$. In Figure~\ref{displacementfits}, we plot the ring displacement $D$ for a Kerr black hole as a function of the inclination angle $i$ for different values of the spin. Triangles denote the data points obtained from our simulation.

We found that the displacement depends primarily on the spin and the disk inclination, but it depends only weakly on the quadrupolar parameter $\epsilon$. The displacement is well approximated by the expression
\begin{equation}
D=2a\sin i(1-0.41\epsilon\sin^2 i),
\label{displacementformula}
\end{equation}
where $a\lesssim0.7M$ for a Kerr black hole whereas $a\leq0.4M$ and $0\leq\epsilon\leq0.5$ for a quasi-Kerr black hole. From our simulations we find a slight deviation for rapidly spinning Kerr black holes with a spin of $a\gtrsim0.7M$ leading to an additional displacement of $0.12a^2\sin i$.

For a Kerr black hole, the dependence of the displacement on the spin and the inclination is reminiscent of the location of the caustics in Kerr spacetime (Rauch \& Blandford 1994; Bozza 2008; see also Bozza 2009 and references therein), and provides an independent check of the nature of the central object. Any deviation from this dependence necessarily implies that the compact object is not a Kerr black hole. For nonzero values of the parameter $\epsilon$, the displacement is only moderately affected.

\begin{figure}[t]
\begin{center}
\includegraphics[width=0.4\textwidth]{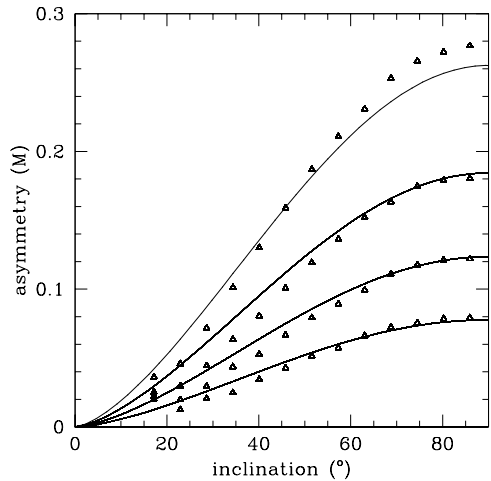}
\includegraphics[width=0.4\textwidth]{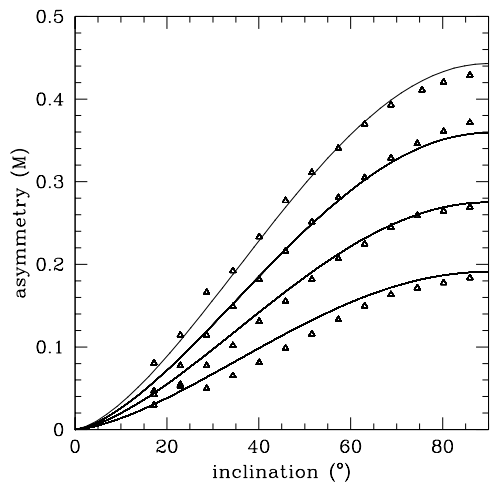}
\end{center}
\caption{The asymmetry of the photon ring as a function of the inclination angle for ({\em top}) a Kerr black hole with values of the spin (top to bottom) $a=0.9M,~0.8M,~0.7M,~0.6M$ and ({\em bottom}) a quasi-Kerr black hole with a spin of $a=0.4M$ and values of the parameter (top to bottom) $\epsilon=0.5,~0.4,~0.3,~0.2$. The solid lines show the functional form (\ref{asymmetryformula}) that describes these simulations.}
\label{asymmetryfits}
\end{figure}

In Figure~\ref{asymmetryfits}, we plot the ring asymmetry as a function of the inclination angle for (top panel) a Kerr black hole with different values of the spin and for (bottom panel) a quasi-Kerr black hole with a spin of $a=0.4M$ and different values of the quadrupolar parameter $\epsilon$. In the case of a Kerr black hole, we find that the asymmetry is negligible for slowly spinning black holes. In the case of a quasi-Kerr black hole, the asymmetry scales linearly with the quadrupolar parameter $\epsilon$. We fit the asymmetry with the expression
\begin{equation}
\frac{A}{M}=\left[0.84\epsilon + 0.36\left(\frac{a}{M}\right)^3\right]\sin^{3/2} i
\label{asymmetryformula}
\end{equation}
which is valid for a Kerr with arbitrary spin and for a quasi-Kerr black hole with values of the spin $0.0\leq a/M \leq0.4$ and of the parameter $0.0\leq\epsilon\leq0.5$, respectively.

The asymmetry depends strongly on the quadrupolar correction parameter $\epsilon$ and the disk inclination $i$ whereas it depends only weakly on the spin as long as the black hole does not spin rapidly. Expression (\ref{asymmetryformula}) implies that the maximum asymmetry of a Kerr black hole is $\simeq0.36M$. In the event that an asymmetry larger than this value is detected, which is already possible for moderate deviations of the quadrupole moment from the value of a Kerr quadrupole moment (c.f., Figure~\ref{asymmetryfits}, bottom panel), then the compact object cannot be a Kerr black hole.

In order to test the no-hair theorem, it is necessary to measure (at least) three multipole moments of the spacetime (Ryan 1995). In the following we argue that a measurement of the diameter of the ring determines the mass of the central object and that for a given disk inclination the displacement and the asymmetry of the ring directly measure the spin $a$ and the parameter $\epsilon$, respectively. Therefore, the degree of asymmetry is a direct measure of the violation of the no-hair theorem.

\begin{figure*}[t]
\begin{center}
\psfig{figure=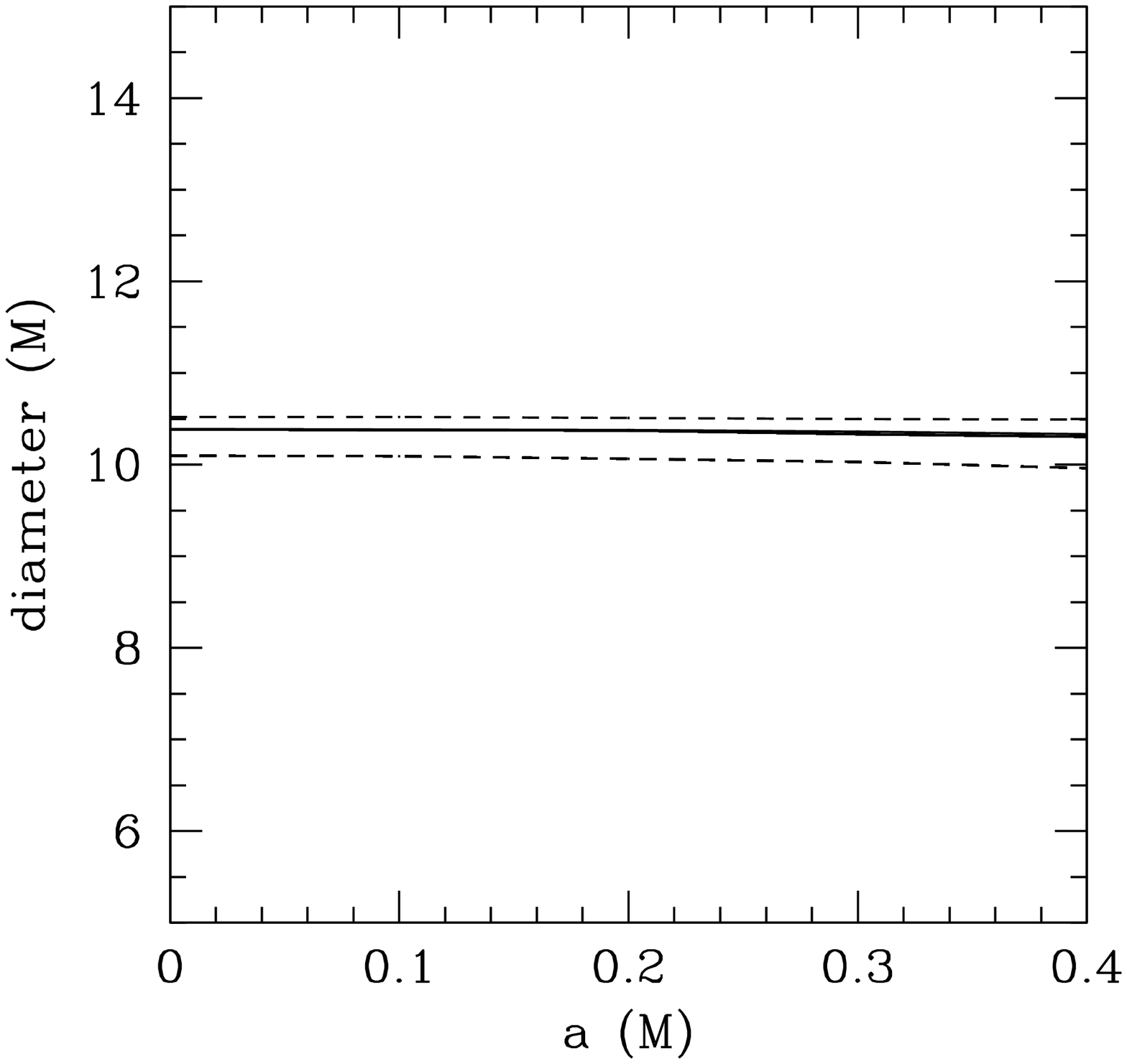,height=2.1in}
\psfig{figure=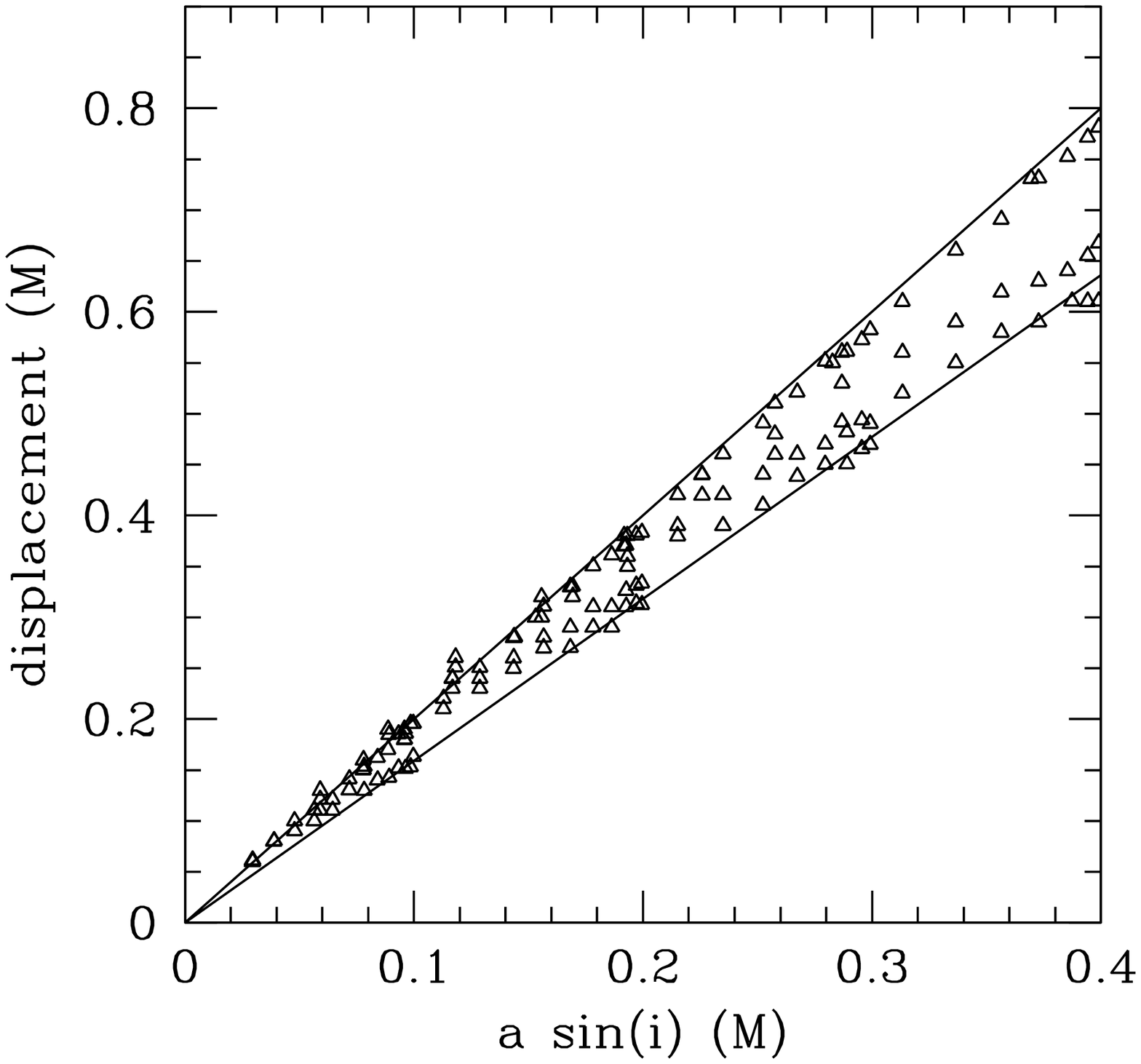,height=2.1in}
\psfig{figure=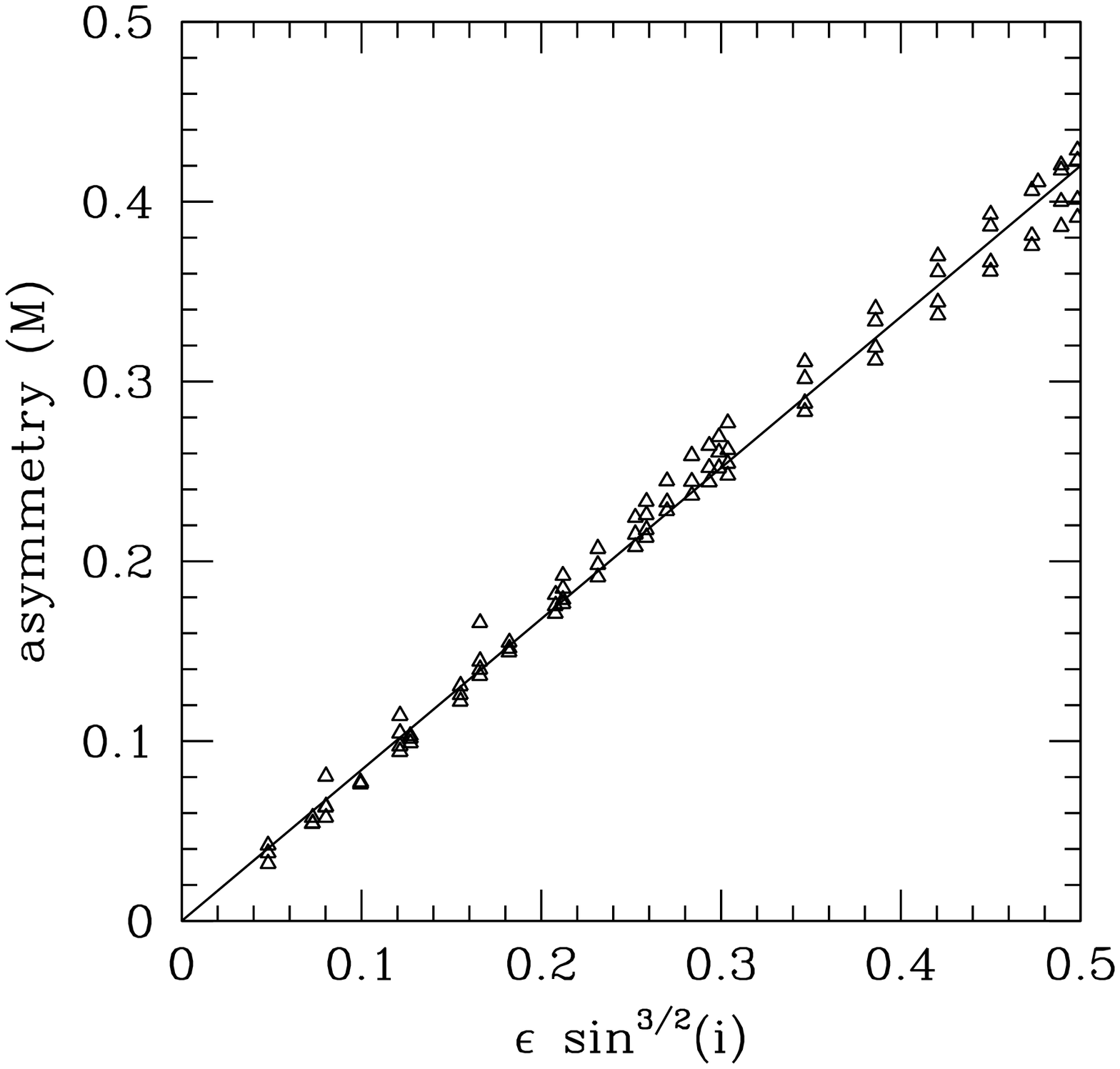,height=2.1in}
\end{center}
\caption{({\em Left:}) The upper and lower limit of the ring diameter versus the spin for inclinations $17^\circ\leq i \leq86^\circ$ for a Kerr black hole (solid lines) and for a quasi-Kerr black hole with a value of the parameter $\epsilon=0.5$ (dashed lines). The diameter is practically independent of the spin and has a constant value of $\simeq10.4M$ for a Kerr black hole. The dependence on the parameter $\epsilon$ and the disk inclination is weak. A measurement of the ring diameter is a direct measure of the mass. ({\em Center:}) The displacement of the photon ring as a function of $a\sin i$ for various values of the parameter $0\leq\epsilon\leq0.5$. The displacement is predominantly determined by the spin and the disk inclination allowing for a direct measurement of the quantity $a\sin i$. ({\em Right:}) The ring asymmetry versus $\epsilon\sin^{3/2} i$ for various inclinations $17^\circ\leq i \leq86^\circ$ and $0.0\leq a/M \leq0.4$. The asymmetry depends only weakly on the spin and hence provides a direct measure of the quantity $\epsilon\sin^{3/2} i$.}
\label{ringstatistics}
\end{figure*}

In Figure~\ref{ringstatistics}, we plot (left panel) the diameter of the photon ring as a function of the spin $a$ for values of the inclination $17^\circ\leq i \leq86^\circ$. Dashed lines correspond to a quasi-Kerr black hole with a value of the parameter $\epsilon=0.5$. The solid line corresponds to a Kerr black hole. In all cases, the diameter is practically independent of the spin $a$. For a Kerr black hole, the diameter also depends only very weakly on the disk inclination and is almost constant for spins in the range $0\leq a/M \leq0.4$ with a value of
\begin{equation}
L=10.4M.
\end{equation}
Even for large values of the spin, the diameter depends only weakly on the inclination angle causing a systematic uncertainty of only $\simeq2\%$ for a value of the spin as large as $a=0.9M$. For a quasi-Kerr black hole, the diameter is affected by the parameter $\epsilon$ and the inclination which leads to a systematic uncertainty of $\simeq5\%$ if $\epsilon=0.5$. Therefore, a measurement of the ring diameter directly measures the mass of the central object.

In Figure~\ref{ringstatistics}, we also plot (center panel) the ring displacement of a quasi-Kerr black hole versus $a\sin i$ for values of the inclination angle $17^\circ\leq i \leq86^\circ$ and of the parameter $0\leq\epsilon\leq0.5$. The solid lines in this figure represent the extreme limits of the fit function given by expression (\ref{displacementformula}) corresponding to $\epsilon=0$ and $\epsilon=0.5$, $\sin i=1$, respectively. Finally, in Figure~\ref{ringstatistics} (right panel), we show the ring asymmetry of quasi-Kerr black holes versus $\epsilon\sin^{3/2} i$ for values of the spin $0\leq a/M \leq0.4$ and of the inclination angle $17^\circ\leq i \leq86^\circ$. The solid line is expression (\ref{asymmetryformula}) evaluated at $a=0.0M$.

Thus, for a given disk inclination, an observation of the displacement and the asymmetry directly leads to a measurement of the spin $a$ and quadrupolar correction parameter $\epsilon$ via the expressions in equations (\ref{displacementformula}) and (\ref{asymmetryformula}), respectively. A measurement of the asymmetry allows for a test of the no-hair theorem. In order to determine the displacement it is necessary to measure the center of mass of the accretion flow around the black hole. An estimate of the location of the center of mass may perhaps be obtained from a (time-averaged) image of the outer edges of the accretion flow.

\vspace{0.7cm}

\section{CONCLUSIONS}

In Paper~I, we proposed a new framework for testing the no-hair
theorem with observations of black holes in the electromagnetic
spectrum. We formulated our tests based on a quasi-Kerr metric
(Glampedakis \& Babak 2006), which deviates smoothly from the Kerr
metric in the quadrupole moment. Since the no-hair theorem admits
exactly two independent multipole moments for a black hole, a
measurement of these three moments will allow us to test the no-hair
theorem.

In this paper, we calculated numerically the mapping between different
locations in the accretion flow around a quasi-Kerr black hole and
positions in the image plane of a distant observer. Our calculations
allowed us to study the potential of using imaging observations of
black holes, that will become available in the near future, in order
to test the no-hair theorem. 

We argued that the expected image of an accretion flow will be
characterized by a bright emission ring generated by light rays that
circle multiple times around the event horizon before emerging towards
the observer. We identified the ring diameter as a direct measure of the mass of the black hole, and we quantified the dependence of the displacement and the asymmetry of the ring on the spin and the quadrupolar parameter as well as on the disk inclination. For a given inclination angle, a measurement of the displacement and the asymmetry directly measures the spin and the quadrupolar parameter of the system, respectively. The asymmetry itself provides a direct measure of the violation of the no-hair theorem.

It is important to emphasize here that only the relative displacement
and asymmetry of the ring (i.e., measured in units of the ring diameter)
and not their absolute values are necessary in inferring the spin and
quadrupole moment of the black-hole spacetime. As a result, the
outcome of such an observation does not depend on the distance to the
black hole, which might not be known accurately. On the other hand, the
angular diameter of the photon ring is proportional to the mass of the 
black hole. This can lead to an accurate measurement of the mass of the 
black hole if the distance is known, or else of the distance to
the black hole, if its mass is known from, e.g., dynamical observations.

Sgr A*, the black hole in the center of the Milky Way, is the ideal
candidate for a test of the no-hair theorem due to its high
brightness, large angular size, and relatively unimpeded observational
accessibility. Recent VLBI observations (Doeleman et al. 2008)
resolved Sgr A* on horizon scales. Incorporating additional baselines
to the VLBI network will lead to the first images of Sgr~A$^{*}$
within the next few years (Fish \& Doeleman 2009). The emission from
Sgr~A$^{*}$ at sub-mm wavelengths is optically thin, the size of the
scattering ellipse is $\lesssim1M$, and the resolution of a VLBI
image at this wavelength is also comparable to $\simeq1M$ (Doeleman
et al.\ 2008; Fish \& Doeleman 2009). The smearing of the images due
to scattering and the finite resolution of the array will, therefore,
not preclude measuring the position and asymmetry of the photon ring
to an accuracy that is adequate in providing a quantitative test of
the no-hair theorem.

Observations of the orbits of stars in the vicinity of the black hole
have provided an independent measurement of its mass (Ghez et
al.\ 2008; Gillessen et al.\ 2009). Perhaps more importantly, the same
observations may lead in the future to an independent measurement of
the spin and orientation of the black hole, as well as to a
complementary test of the no-hair theorem (Will 2008; Merritt et
al.\ 2010).

\acknowledgements

We thank A.\ Broderick, S.\ Doeleman, V.\ Fish, K.~ Glampedakis, S.\ Hughes, A.\ Loeb, D.\ Marrone, F.\ \"Ozel, J. \ Steiner, and R. \ Takahashi for many useful discussions. DP thanks the Institute for Theory and Computations at Harvard University for their hospitality. This work was supported by the NSF CAREER award NSF 0746549.

\appendix

\section{PHOTON INITIAL CONDITIONS}
\label{inicond}

For the geometry in Figure \ref{geometry}, the initial conditions of
the photons in the image plane of a distant observer are better
expressed in spherical coordinates. The image plane is located at a
distance $d$ from the black hole at an inclination angle $i$. Since
the observer is far away from the black hole, we can express the
initial conditions using standard Euclidean geometry.

First, we transform the image plane coordinates $(x',y',z')$ into
Cartesian coordinates $(x,y,z)$ centered at the black hole. We find
\begin{eqnarray}
x&=&-y'\cos i + z'\sin i + d\sin i\;,\nonumber\\
y&=&x'\;,\nonumber\\
z&=&y'\sin i + z'\cos i + d\cos i\;.
\end{eqnarray}
We then convert the Cartesian black-hole coordinates into spherical
coordinates by the usual transformation
\[
r=\sqrt{x^2+y^2+z^2},
\]
\[
\theta=\arccos\frac{z}{r},
\]
\begin{equation}
\phi=\arctan\frac{y}{x}.
\end{equation}

The initial conditions for a photon on the image plane at $(x',y')$
with uniform initial momentum $\vec{k}_0=-k_0\hat{z}'$ are then
given by
\[
r_0=\sqrt{x'^2+y'^2+d^2},
\]
\[
\theta_0=\arccos{\frac{y'\sin i+d\cos i}{r_0}},
\]
\begin{equation}
\phi_0=\arctan{\frac{x'}{d\sin i-y'\cos i}},
\end{equation}
\noindent
and
\[
k_{r_0}=-\frac{d}{r_0}k_0,
\]
\[
k_{\rm \theta_0}=\frac{\cos i - (y'\sin i + d\cos i)\frac{d}{r_0^2}}{\sqrt{ x'^2 + \left(d\sin i - y'\cos i\right)^2}}k_0,
\]
\begin{equation}
k_{\rm \phi_0}=\frac{x'\sin i}{x'^2+ (d\sin i - y'\cos i)^2}k_0,
\end{equation}
\noindent
where $r_0$, $\theta_0$, and $\phi_0$ are the initial coordinates of a
given photon, and $k_{\rm r_0}$, $k_{\rm \theta_0}$, and $k_{\rm
  \phi_0}$ are the components of its initial 3-momentum. The time
component of the 4-momentum is calculated from the 3-momentum $\vec{k}_0$ so
that its norm vanishes.

%\section*{References}

\end{document}